\documentstyle[preprint,aps,epsf,floats]{revtex}
\tighten

\def\si{{}^1\kern-.14em S_0}
\def\siii{{}^3\kern-.14em S_1}
\def\diii{{}^3\kern-.14em D_1}
\def\bfp{{\bf p}}
\def\bfq{{\bf q}}
\def\pds{$PDS$\ }
\def\fm{{\rm\ fm}}
\def\MeV{{\rm\ MeV}}
\def\CA{{\cal A}}

\begin{document}
\preprint{\vbox{
\hbox{DOE/ER/40561-352-INT97-00-189}
\hbox{NT@UW-98-05}
\hbox{CALT-68-2155} }}
\title{A New Expansion for  Nucleon-Nucleon Interactions}
\author{David B. Kaplan }
\address{
Institute for Nuclear Theory 351550, University of Washington, Seattle, WA
98195-1550, USA}
\author{Martin  J. Savage }
\address{
Department of Physics 351560, University of Washington,  
Seattle, WA   98195-1560, USA }
\author{Mark B. Wise }
\address{
California Institute of Technology, Pasadena, CA  91125, USA }
\maketitle

\begin{abstract} 
We introduce a new and well
defined power counting for the effective field theory describing
 nucleon-nucleon interactions.  Because of the large $NN$ scattering
lengths it differs from other applications of chiral perturbation theory and
is facilitated by introducing an unusual subtraction scheme and 
renormalization group analysis. Calculation to 
subleading order in the expansion can be done analytically, 
and we present the results for both the 
$\si$ and $\siii-\diii$ channels.
\end{abstract}

\bigskip
\vskip 8.0cm
\leftline{January 1998}

\vfill\eject

Effective field theories are standard  for dealing with strong
interaction physics in the nonperturbative regime.  The idea is to
construct the most general Lagrangian consistent with the 
symmetries of quantum
chromodynamics out of fields that create and destroy the relevant 
degrees of freedom.  To have predictive power an expansion
scheme must be found that limits the number of terms that occur in the
effective Lagragian.  In the case of the effective Lagrangian for pion self
interactions  the expansion is in powers of the typical momentum
$p \sim m_\pi$ divided by a typical hadronic scale, of order the rho mass.  
At leading order, tree diagrams dominate with either two derivatives
or one insertion of the light quark
mass matrix at the vertex.  
At next order one includes both four 
derivative operators at tree 
level, as well as one loop diagrams with the leading operators at the vertices,
and so forth.
This perturbative expansion is known as chiral perturbation theory, and 
has also been applied to nucleon-pion interactions.  However, attempts to apply
chiral perturbation theory to nuclear physics, where nucleon-nucleon
interactions are of primary interest
\cite{We90a,KoMany,Parka,KSWa,CoKoM,DBK,cohena,Fria,Sa96,LMa,GPLa,Wisea,Adhik,RBMa,Bvk,Parkb}, 
have encountered difficulties, stemming from the large scattering lengths in the $\si$ 
and $\siii$ nucleon-nucleon scattering amplitudes \cite{We90a,KSWa}.  The large
scattering lengths generally imply finely tuned cancellations between graphs in the 
effective theory, with the result that a consistent expansion has so far proven elusive.

In this letter we introduce a new method for applying effective field theory to
nuclear physics.  Our method uses dimensional regularization with a novel
subtraction scheme which permits a consistent  expansion procedure yielding
analytic expressions for  nucleon-nucleon scattering amplitudes 
\footnote{Dimensional regularization is 
the most convenient regularization scheme, as it preserves chiral symmetry and gauge 
symmetry; it also preserves Galilean invariance, which makes the Feynman integrals 
relatively simple 
to evaluate.  However, physical results do not depend upon the 
choice of regulator.}.  We apply it
to  scattering amplitudes in both the $\si$ and $\siii-\diii$ 
channels
and compare with the results of the Nijmegen partial wave analysis \cite{Nijmegen}.

We begin by considering nucleon-nucleon scattering in the $\si$
channel in a nonrelativistic effective field theory with only nucleon fields,
as is appropriate for momentum much less than the pion mass.
The tree level amplitude is
\begin{eqnarray}\label{1}
iA_{tree} &=& - i (\mu/2)^{4-D} \sum_{n = 0}^\infty C_{2n} (\mu) p^{2n}\nonumber \\
&=& -i (\mu/2)^{4-D} C(p^2, \mu)
\ \ \ ,
\end{eqnarray}
where $p$ is the magnitude of the nucleon momentum in the center of mass frame;
the total center of mass energy is 
$E = p^2/M+\ldots$, where $M$ is the nucleon mass.  In
eq.~(\ref{1}) $C_{2n} (\mu)$ is a linear combination of coefficients of four
nucleon operators with $2n$ spatial derivatives, $\mu$ is the subtraction
point and $D$ is the dimension of spacetime.  In order to calculate the full
amplitude we must sum a bubble chain of Feynman diagrams with the tree
amplitude giving the vertices.  The integral required is
\begin{eqnarray}
I_n  &=&  - i (\mu/2)^{4-D} \int {d^D q\over (2\pi)^D} \bfq^{2n}  {i\over (E/2 - q^0 -
\bfq^2/2M + i\varepsilon)}\cdot {i\over (E/2 + q^0 - \bfq^2/2M +
i\varepsilon)}
\nonumber\\
&=& - M (ME)^n (- ME - i\varepsilon)^{D - 3\over 2} \Gamma \left({3 - D\over 2}\right)
{(\mu/2)^{4-D}\over (4\pi)^{(D - 1)/2}}
\ .
\end{eqnarray}
This amplitude has no pole at $D = 4$, but it does have a pole at $D=3$
resulting from what would be a linear ultraviolet divergence in the integration
if a momentum cutoff were used.  The residue of the pole
at $D = 3$ equals $M(ME)^n \mu/4\pi$.  In the minimal subtraction scheme, $MS$, counter terms
are added which subtract the poles at $D = 4$.  Here we use a new subtraction
procedure which we call power divergence subtraction, \pds, where the poles at
$D = 4$ and $D = 3$ are both subtracted by counter terms.  In the case of $I_n$
this subtraction amounts to adding the counter term contribution
\begin{equation}
\delta I_n = -{M(ME)^n \mu\over 4\pi (D-3)},
\end{equation}
and the subtracted integral is
\begin{equation}
I_n^{PDS} = I_n + \delta I_n = - (ME)^n \left({M\over 4\pi}\right) (\mu + ip).
\end{equation}
The choice $\mu = 0$ corresponds to minimal subtraction.  Using the \pds
subtraction scheme the full amplitude, including loop effects 
is \footnote{Some
recent papers have criticized the applicability of dimensional regularization
to $NN$ scattering\cite{cohena,GPLa,RBMa}.  
Since $C$ is an arbitrary polynomial in $p^2$, over a
range in $p$, eq. (6) gives the most general form for $S$ consistent with
unitarity.  Clearly dimensional regularization can be used for $NN$
scattering.
Relativistic corrections have been neglected in eq. (5).
They can be included perturbatively by inserting higher derivative one-body
operators and corrections to the energy-momentum relation for external legs.}
\begin{equation}\label{6}
i{\cal A} = {-i C(p^2, \mu)\over 1 + MC(p^2, \mu) (\mu + ip)/4\pi}.
\end{equation}

The scattering amplitude ${\cal A}$ is related to the $S$-matrix by
\begin{equation}
S-1 = e^{2i\delta}-1 = i \left({pM\over 2\pi}\right) {\cal A}
\ ,
\end{equation}
where $\delta$ is the phase shift.  It is convenient to consider
the quantity
\begin{equation}\label{8}
p \cot\delta = ip + {4\pi\over M{\cal A}} = - {4\pi\over MC(p^2, \mu)} - \mu\ ,
\end{equation}
which has an expansion in $p^2$ of the form
\begin{equation}\label{9}
p\cot\delta = - {1\over a} + {1\over 2} r_0 p^2 + \ldots,
\end{equation}
where $a$ is called the scattering length and $r_0$ is the effective range.  If
there is a state that is very near zero binding energy then $|a|$ will be very
large, which is the case in the $\si$ channel where $a = - 23.714 \pm 0.013 \fm$ and 
$r_0 = 2.73 \pm 0.03 \fm$.  The effective range $r_0$ and
the higher coefficients in the momentum expansion are bounded by the range of
the nuclear potential.

The coefficients $C_{2n} (\mu)$ are determined by $a, r_0, \ldots $ etc.  Their
$\mu$ dependence cancels the explicit $\mu$ dependence in eq.~(\ref{6}) so that
the physical scattering amplitude ${\cal A}$ is $\mu$ independent.  The $\mu$
independence of ${\cal A}$ implies a set of renormalization group equations for
the $C_{2n}$'s.
\begin{equation}
\mu {d\over d\mu} C_{2n} = \beta_{2n}\ ,
\end{equation}
where, for example,
\begin{equation}
\beta_0 = {M\over 4\pi} C_0^2 \mu  \qquad \qquad ~{\rm and}~ \qquad \beta_2 =
{M\over 2\pi} C_0 C_2 \mu
\ ,
\end{equation}
are the exact beta-functions for $C_0$ and $C_2$.
Solving these equations with boundary conditions supplied by eqs.~(\ref{8},\ref{9}) 
yields
\begin{equation}\label{11}
C_0 (\mu) = {4\pi\over M} \left({1\over -\mu + 1/a}\right) \qquad \qquad ~{\rm
and}~ \qquad 
C_2 (\mu) = {2\pi\over M} \left( {1\over -\mu + 1/a}\right)^2 r_0.
\end{equation}
In a system with a short range interaction characterized by the 
momentum scale $\Lambda$, 
the natural size for the parameters $a,r_0,\ldots$ is set by the scale
$\Lambda$. When $a$ is of natural size (i.e. in magnitude 
of order $1/\Lambda$), 
 a simple power counting like that used in chiral 
perturbation theory for
pion self interactions is possible.  
It is then convenient to take $\mu = 0$ which corresponds to the $MS$ scheme.
The amplitude ${\cal A}$ can be
expanded in powers of  $p/\Lambda$ with each loop contributing a
factor of $(M p/4\pi)$ and each vertex a factor of, 
$C_{2n}p^{2n}\sim (4\pi p^{2n})/(M\Lambda^{2n+1})$. 
Consequently the leading contribution of order $p^0$ comes from
$C_0$ at tree level.  At order $p^1$ a one loop diagram with two $C_0$'s
contributes.  At order $p^2$ one must include a two loop diagram with $C_0$ at
each vertex and a tree level diagram with a vertex proportional to $C_2$, etc.

In contrast, for the special case where $|a|$ is very large and $\mu=0$, 
the coefficients are very large, $C_{2n} \sim (4 \pi a^{n+1})/(M\Lambda^n)$, 
and the  momentum expansion used for $|a| \sim 1/\Lambda$ is  
valid only over a small  momentum range.  
This difficulty is cured by taking $\mu$ nonzero in which case  the
coefficients $C_{2n}$ need not be large.
As $|a\mu| \rightarrow \infty$ one finds, 
$C_{2n} \sim 4\pi/(M\Lambda^n \mu^{n+1})$.
If one were to set $\mu=\Lambda$, this would be 
similar to the result of \cite{GPLa} where a momentum
cut off $\Lambda$ is used.

The $\mu$ dependence induced by the additional finite subtraction in the
\pds scheme
 makes it possible to establish a power counting that allows a systematic
expansion of
the scattering amplitude over a much larger range of momentum.  Choosing $\mu
\sim p$, all Feynman diagrams with $C_0$'s at each vertex give a contribution of
the same  order.  This occurs because $C_0$ is of order $p^{-1}$ while each loop
integration is of order $p$.  With $\mu \sim p$ the coefficients $C_{2n}$ are
of order $p^{-(n + 1)}$ and $C_{2n} p^{2n} \sim p^{n-1}$.  Consequently only
$C_0$ must be treated to all orders.  The other coefficients can be treated
perturbatively.  The scattering amplitude thus has an expansion of the form
\begin{equation}
{\cal A} = {\cal A}_{-1} +  {\cal A}_0 + {\cal A}_1 +\ldots
\end{equation}
where ${\cal A}_q$ is of order $p^q$.  The leading amplitude is given by the sum of 
loop graphs with $C_0$ vertices, 
while subleading contributions correspond to perturbative insertions 
of the momentum dependent interactions, dressed to all orders by $C_0$.  The first 
two terms in the expansion of $\CA$ are
\begin{equation}\label{14}
{\cal A}_{-1} = {- C_0\over \left[1 + {C_0 M\over 4\pi} (\mu + ip)\right]}
\qquad {\rm and} \qquad {\cal A}_0 = { - C_2 p^2\over \left[1 + {C_0 M\over 4\pi}
(\mu + ip)\right]^2}\ ,
\end{equation}
which arise from the graphs shown in Fig.~1.
\begin{figure}[t]
\centerline{\epsfysize=2.5 in \epsfbox{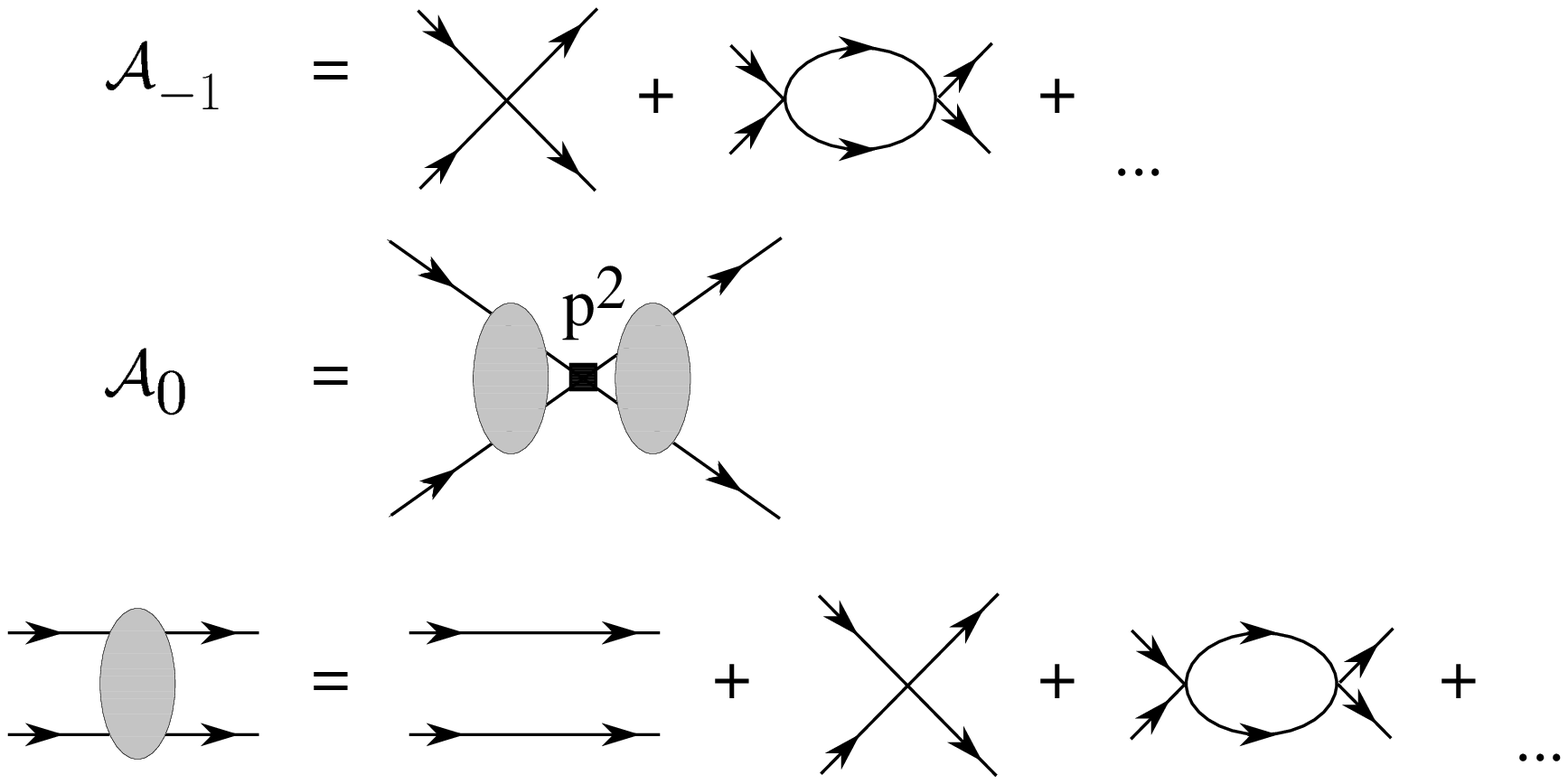}}
\noindent
Fig~1. {\it The Feynman graphs contributing to $\CA_{-1}$ and  $\CA_{0}$.  
Each of the resummed vertices 
is a factor of  $C_0$, while the vertex marked with a square is a factor of $C_2p^2$.
Summing the graphs yield the  expressions in eq.~(\ref{14}).}
\vskip .2in
\end{figure}

As we are interested in momenta $p \sim m_\pi$ explicit 
pion fields must be included in the effective Lagrangian.  The pion
couplings are determined from the usual chiral Lagrangian for pion-nucleon
interactions.  
Since we are considering nucleon-nucleon interactions
in a theory where the nucleons are treated nonrelativistically, it is convenient
to consider separately  the pion interaction which gives rise to a static $N N$ 
potential and the radiation pion field responsible for the remaining 
effects.  This
is similar to the treatment of the photon field in nonrelativistic QED
\cite{LMa,NRQCD}. 
The resulting effective field theory 
 amounts to expanding the dependence of nucleon propagators on the
3-momentum of the pion fields when the energy loop integrations get a
contribution from the pole in the pion propagator.  In the $^1S_0$
channel the
momentum space potential from one pion exchange is
\begin{equation}
V_\pi (\bfp, \bfp') = - {g_A^2\over 2f^2} \left({m_\pi^2\over \bfq^2 +
m_\pi^2} -1 \right),
\end{equation}
where $\bfq = \bfp' - \bfp$,  $g_A = 1.25$ and $f = 132\MeV$ is the pion
decay constant.  It gives an order $p^0$, spatially extended, 
nucleon-nucleon interaction
since we are treating $p \sim m_\pi$.  Hence including pions does not change
the order ${\cal A}_{-1}$ term, and the graphs summed at leading order are still 
those in Fig.~1.  For the
order $p^0$ contribution there are additional
terms from pion exchange: aside from the contribution to $\CA_0$ shown in Fig.~1, 
one must also include the exchange of one potential pion dressed by $C_0$ to all orders
(radiation pions do not contribute at this order).

The expression for ${\cal A}_0$ is conveniently expressed as the sum of the five 
graphs shown in Fig.~2; $\CA_0^{(I)}$ is the
contribution to $\CA_0$ in eq.~(\ref{14}), while
\begin{eqnarray}\label{17}
 {\cal A}_0^{(II)} &=&  \left({g_A^2\over 2f^2}\right) \left(-1 + {m_\pi^2\over
4p^2} \ln \left( 1 + {4p^2\over m_\pi^2}\right)\right)\ ,\nonumber\\
 {\cal A}_{0}^{(III)} &=& {g_A^2\over f^2} \left( {m_\pi M{\cal A}_{-1}\over 4\pi}
\right) \Bigg( - {(\mu + ip)\over m_\pi}
+ {m_\pi\over 2p} \left[\tan^{-1} \left({2p\over m_\pi}\right) + {i\over 2} \ln
\left(1+ {4p^2\over m_\pi^2} \right)\right]\Bigg)\ ,\nonumber\\
{\cal A}_0^{(IV)} &=& {g_A^2\over 2f^2} \left({m_\pi M{\cal A}_{-1}\over
4\pi}\right)^2 \Bigg(-\left({\mu + ip\over m_\pi}\right)^2
+ \left[ i\tan^{-1} \left({2p\over m_\pi}\right) - {1\over 2} \ln
\left({m_\pi^2 + 4p^2\over\mu^2}\right) + 1\right]\Bigg)\ ,\nonumber\\
{\cal A}_0^{(V)} &=& - {D_2 m_\pi^2\over\left[1 + {C_0 M\over 4\pi} (\mu +
ip)\right]^2}\ .
\end{eqnarray}
Note that $\CA_0^{(IV)}$  has a logarithmic dependence on $\mu$.
Since $m_\pi^2$ 
is proportional to the
light quark masses we are required to include the vertex in $\CA_0^{(V)}$ 
proportional to $D_2 m_\pi^2$
\space\footnote{The presence of such a divergence \cite{KSWa} 
indicates that the power
counting suggested by Weinberg \cite{We90a} is not consistent.  
The influence of this
operator is not subdominant to one pion exchange when $C_0$ is treated to all
orders.  Its presence is needed to get a subtraction point independent
amplitude.}. 
\begin{figure}[t]
\centerline{\epsfysize=2.5 in \epsfbox{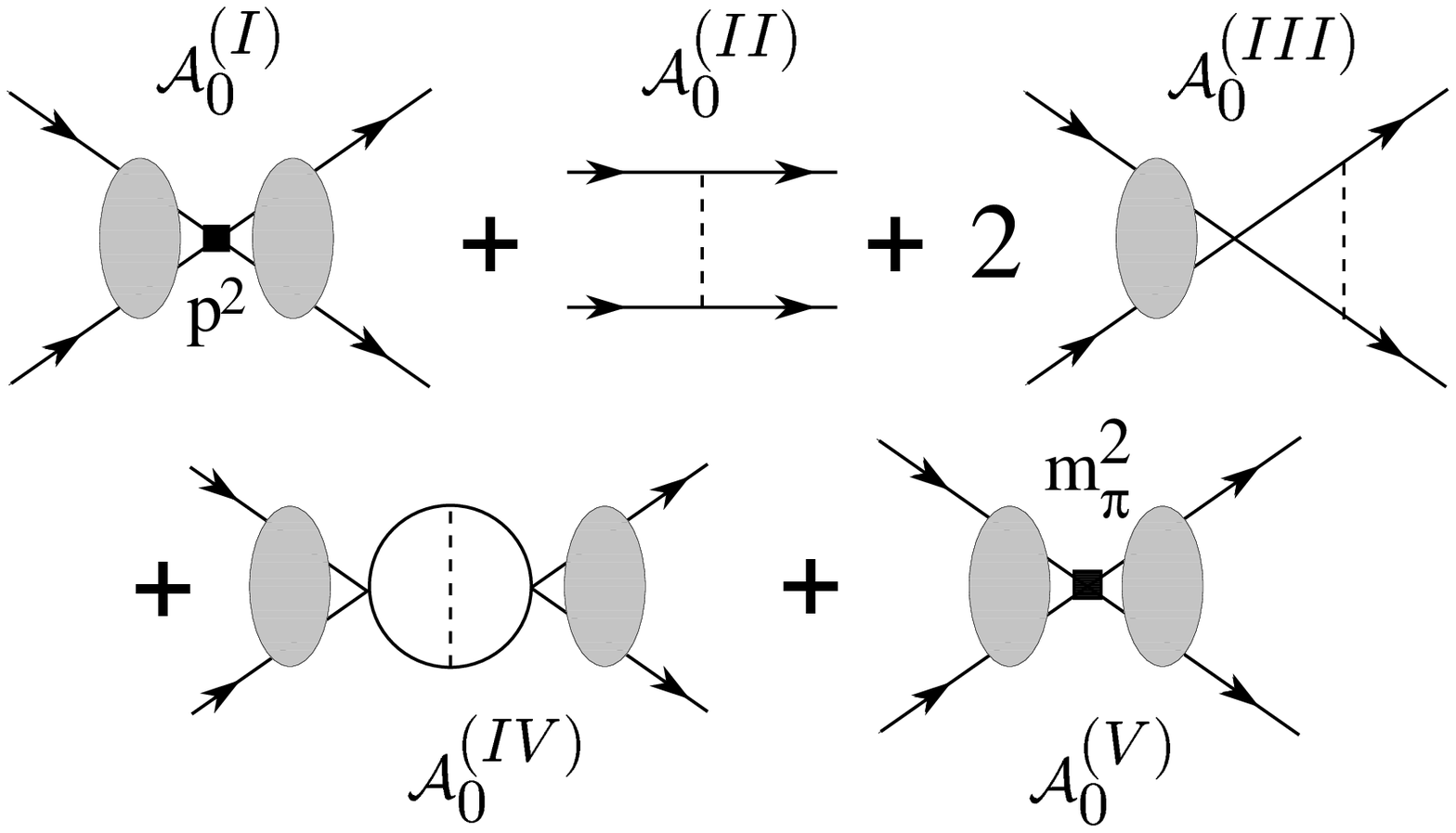}}
\noindent
Fig~2. {\it  The five sub-graphs contributing to $\CA_0$ computed in eq.~(\ref{17}). 
The gray blob is defined in Fig.~1, while the dashed line is the exchange of a potential pion.}
\vskip .2in
\end{figure}
We have absorbed into ${\cal A}_0^{(V)}$  some  of the finite, 
$\mu$-independent, part of ${\cal A}_0^{(IV)}$
that arises in $PDS$ (e.g. the part involving Euler's
constant and a logarithm of $4\pi$).  
In our power counting $p,\mu$ and $m_\pi$
are considered to be of the same order.  Comparing ${\cal A}_{-1}$ with ${\cal
A}_0$ indicates that the expansion parameter is $p/\Lambda_{NN}$ 
where
\begin{equation}
\Lambda_{NN} = (8 \pi f^2/g_A^2 M) \sim 300\MeV.
\end{equation}
At leading order in our expansion, for large scattering length, 
$C_0({m_\pi})= -4\pi/(Mm_{\pi})=-3.7\fm^2$. 
Our power counting leads us to expect that  $|C_2(m_{\pi})|$ and 
$|D_2(m_{\pi})|$ are of order $4\pi /(M\Lambda_{NN}m_{\pi}^2)\sim 3.5\fm^4$.

The inclusion of the pions changes the
renormalization group equations for $C_0$ and $C_2$ from what is
given in eq.~(\ref{11}).  
To the order we are working the new beta-functions are
\begin{equation}
\beta_0 = {M\over 4\pi}\left(C_0^2 + {C_0 g_A^2\over f^2}\right)\mu ~\qquad
{\rm
and} \qquad~ \beta_2 =  {M\over 2\pi} C_0 C_2 \mu
\ .
\end{equation}
In addition the constant $D_2 (\mu)$ satisfies the renormalization group
equation
\begin{equation}
\mu {d\over d\mu} D_2 =  {M\over 2\pi} D_2 C_0\mu + {M^2 g_{\cal A}^2\over
32\pi^2 f^2} C_0^2
\ .
\end{equation}
The solutions to these renormalization group equations are the 
running coupling constants $C_0(\mu)$, $C_2(\mu)$ and $D_2(\mu)$.
When
expressed in terms of the running couplings, the full amplitude $\CA$ is independent 
of $\mu$; however, as in perturbative QCD, our expression for $\CA_{-1}+\CA_0$ 
will be $\mu$ independent only up to the order we are working, $p^0$, with residual 
$\mu$ dependence  canceled by higher order terms in the expansion of $\CA$. 
The free parameters can be taken to be the value of the running coupling constants
at a particular scale, which we choose to be $\mu=m_\pi$.

All that remains is to fit our free parameters 
$C_0(m_\pi)$, $C_2(m_\pi)$ and $D_2(m_\pi)$
to data.  
We compute the phase shift 
$\delta$  from the exact expression
\begin{eqnarray}
\delta & = & 
{1\over 2i} \ln\left( 1 + i {Mp\over 2\pi}\CA \right)
\ ,
\end{eqnarray}
expanding both sides to a given order with 
$\delta = \left( \delta_{(0)} + \delta_{(1)} + \ldots
\right) $.
For example, to subleading order,
\begin{eqnarray}
\delta_{(0)} & = & 
{1\over 2i} \ln\left( 1 + i {Mp\over 2\pi}\CA_{-1} \right)
\ ,\qquad
\delta_{(1)} = {M p \over 4 \pi} \left( {\CA_0\over 1 + i {Mp\over 2\pi}\CA_{-1} }
\right)
\ .
\end{eqnarray}

We could determine the free parameters analytically by requiring that 
the theory exactly reproduce effective range theory at low momentum.  
If we do this, our
result reproduces the data very well up to center of mass 
momentum $p\sim 150\MeV$.  
However,  better  agreement with the data at large momenta 
results from a fit to the phase shift 
from the Nijmegen partial wave  analysis \cite{Nijmegen} 
optimized over the momentum range $p\le 200 \MeV$.  
The result of this procedure are the values
\begin{equation}\label{numfit}
C_0(m_{\pi})=-3.34\fm^2\ ,\qquad
D_2(m_{\pi})=-0.42\fm^4\ ,\qquad
C_2(m_{\pi})=3.24\fm^4\ .
\end{equation}
and the phase shift plotted in Fig.~3. (The individual values of
$C_0$ and $D_2$ are not particularly meaningful since up to terms higher
order in our expansion the amplitude can be written in terms of the linear
combination $C_0+D_2m_{\pi}^2$.)
As is apparent from Fig.~3, the agreement of the phase shift with data 
is excellent at quite large values of $p$. Furthermore,
the coupling $C_0(m_{\pi})$
is close to its leading order value (in the limit of large scattering length),
$-3.7\fm^2$ and $C_2(m_{\pi})$ is at the expected size, suggesting that
our expansion is valid in this channel. However, for
$p>100\MeV$ the magnitude of the ratio
${\cal A}_0/{\cal A}_{-1}$ is greater than $\sim 0.5$ and it is
difficult to justify the approximations we have made, e.g.
neglecting terms suppressed by  $({\cal A}_0/{\cal A}_{-1})^2$.

We have performed a similar analysis in the coupled $\siii-\diii$ NN
scattering channels. 
In this channel the amplitude ${\cal A} $ is a $2\times 2$ matrix with
elements ${\cal A}_{L,L^{\prime}}$. The S-matrix
in this channel is usually expressed in terms of two phase shifts,
$\delta_{0}$,
$\delta_{2}$ and a mixing angle $\varepsilon_1$,
\begin{figure}[t]
\centerline{\epsfysize=4.5 in \epsfbox{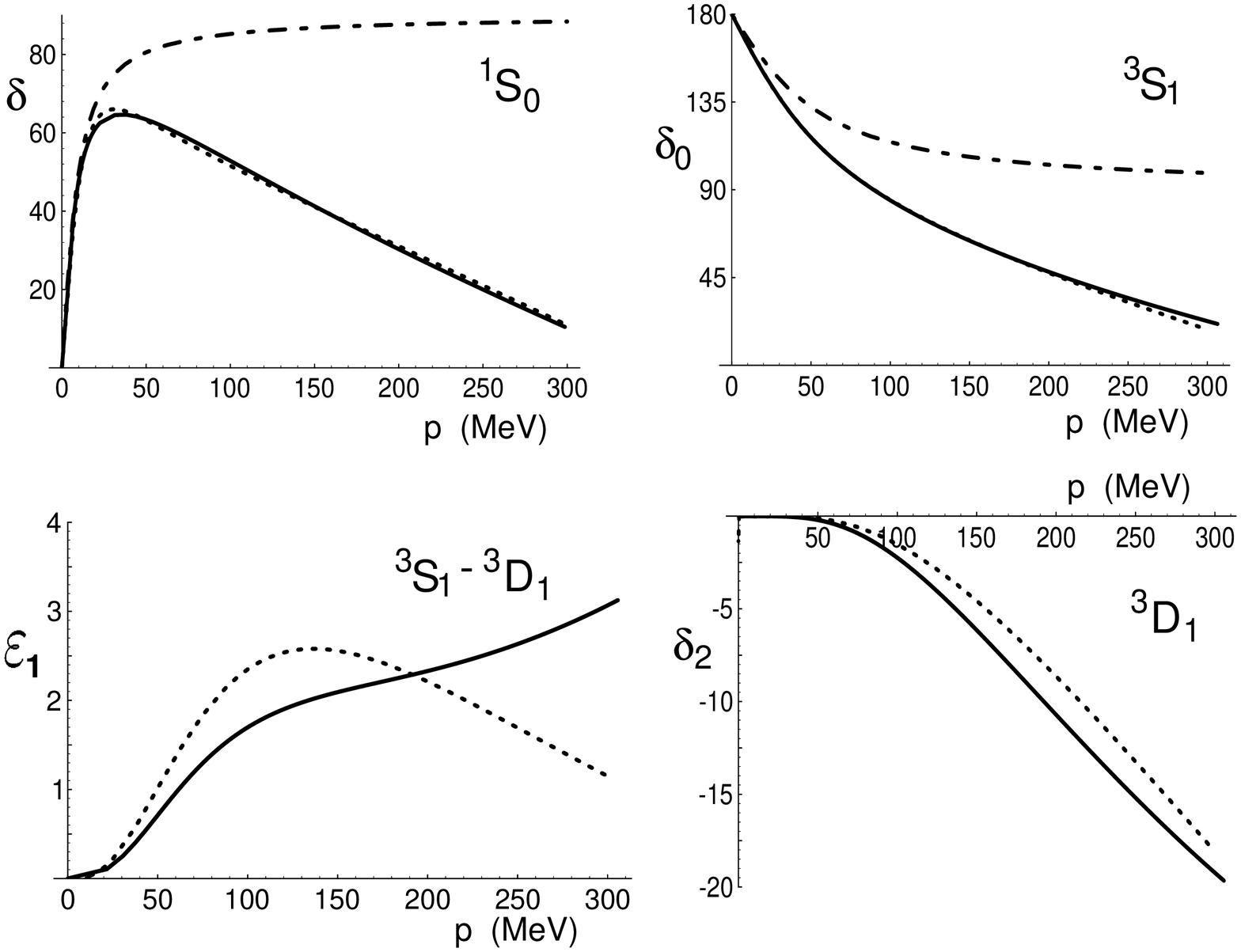}}
\noindent
Fig~3. {\it The phase shifts $\delta$ in the $\si$ channel, and $\delta_{0}$, 
$\varepsilon_1$,  and $\delta_{2}$ in the $\siii-\diii$ coupled channels, plotted
in degrees versus center of mass momentum $p$.  
The dot-dashed line represents the  leading $p^{-1}$  order calculation for 
$\delta$ and $\delta_{0}$; at this order one finds  $\varepsilon_1=
\delta_{2}=0$. The dashed lines are the results from the order $p^0$ 
calculation.   Solid lines are results
from the Nijmegen partial wave analysis 
of scattering data \cite{Nijmegen}.  The fits involve three parameters in the 
$\si$ channel and three parameters in the $\siii-\diii$ coupled channels.}
\vskip .2in
\end{figure}
\begin{equation}
S-1=  i{pM \over 2\pi}{\cal A}=\left( \begin{array}{ll}
e^{2i\delta_{0}} \cos 2\varepsilon_{1} -1 & ie^{i(\delta_{0} + \delta_{2})}
\sin 2\varepsilon_1\\
ie^{i(\delta_{0} + \delta_{2})} \sin 2\varepsilon_1 & e^{2i\delta_{2}} \cos
2\varepsilon_1 -1\end{array} \right) 
\end{equation}
The scattering length is large in the $\siii$ channel 
($a = 5.423\pm 0.005\ {\rm fm}$)
and the power counting is 
analogous to that of the $\si$ channel. At
leading order, $p^{-1}$, ${\cal A}$ is expressed in terms of a single parameter
that is a linear combination of coefficients of four-nucleon operators with no
derivatives. Thus ${\cal A}_{02}={\cal A}_{20}=
{\cal A}_{22}=0$, which implies that $\varepsilon_1 = \delta_{2}=0$. 
At next order, $p^0$, there are two new
parameters analogous to $C_2$ and $D_2$ that occur in the expression for
${\cal A}_{00}$. 
In a future publication \cite{KSW3}\  analytic formulae for the
elements
${\cal A}_{LL^{\prime}}$ will be given to order $p^0$. Results of fitting
the three parameters that occur in ${\cal A}_{00}$ to the Nijmegen partial wave analysis
\cite{Nijmegen}\  are presented in Fig.~3.  Note that we have worked to subleading order
in calculating $\delta_{0}$, but only leading order in $\varepsilon_1$ and 
$\delta_{2}$, which explains the relative quality of the fits.
After fitting $\delta_{0}$ there are no free parameters in the 
predictions for $\varepsilon_1$ and $\delta_{2}$ at this order.

In conclusion, we have shown how to describe $NN$ scattering in 
a dimensionally regulated theory with a consistent expansion, and have 
demonstrated the failure of Weinberg's power counting for these systems.
The subtraction scheme we have introduced is useful for treating 
finely tuned theories with power-law divergences such as the standard model
with a light Higgs, and the Nambu--Jona-Lasinio model, 
as well as any nonrelativistic theory with 
a large scattering length arising from a short-range interaction.
In a future publication we will present other applications 
of the methods developed here, including a discussion of other 
partial waves, inelastic processes,  properties of the
deuteron, and the $N$-body 
problem\cite{KSW3}.

\bigskip\bigskip

We would like to thank P. Bedaque, A. Bulgac and U. van Kolck  for
useful discussions.
This work supported in part by the U.S. Dept. of Energy under
Grants No. DOE-ER-40561,  DE-FG03-97ER4014, and DE-FG03-92-ER40701.

\end{document}